\documentclass[aps,prd,12pt,notitlepage,final,oneside,onecolumn,nobibnotes,
nofootinbib,superscriptaddress,noshowpacs,centertags,amsmath,amssymb,showkeys]{revtex4-2}

\usepackage[utf8]{inputenc}
\usepackage[english]{babel}
\usepackage{graphics}
\usepackage{amsmath}
\usepackage{amssymb}
\usepackage{mathtools}
\usepackage{hyperref}

\newcommand{\diag}{\text{diag}}
\newcommand{\Tp}{\text{T}}
\newcommand{\operP}{\mathcal{P}}

\begin{document}

\title{Fermion propagator diagonalization and eigenvalue problem}

\author{D.\,A. Dolzhikov}%
\email{loldololo@gmail.com}%
\affiliation{Joint Institute for Nuclear Research, 141980, Dubna, Russia}

\author{A.\,E. Kaloshin}%
\email{kaloshin@physdep.isu.ru}%
\affiliation{Irkutsk State University, 664003, Irkutsk, Russia}%
\affiliation{Joint Institute for Nuclear Research, 141980, Dubna, Russia}

\author{V.\,P. Lomov}%
\email{lomov.vl@icc.ru}%
\affiliation{%
  Institute for System Dynamics and Control Theory, RAS, 664033, Irkutsk,
  Russia}%
\affiliation{Irkutsk State University, 664003, Irkutsk, Russia}%

\begin{abstract}
  We discuss diagonalization of propagator for mixing fermions system based on
  the eigenvalue problem. The similarity transformation converting matrix
  propagator into diagonal form is obtained. The suggested diagonalization has
  simple algebraic properties for on-shell fermions and can be used in
  renormalization of fermion mixing matrix.
\end{abstract}


\maketitle


\section{Introduction}
\label{sec:intro}

The problem of fermion mixing is widely discussed in last decades, especially in
connection with neutrino experiments. It is generally recognized that the most
adequate description of neutrino oscillations should be based on the quantum
field theory (QFT) methods. There exists a lot of papers on the topic, touching
on various aspects of the problem, see \cite{Gri96,Giunti:2002xg, Beu03,
  Akh10,Dvo11,Martone:2011kh,BahaBalantekin:2018ppj,Blasone:2019rxl,
  Grimus:2019hlq, Naumov:2020yyv} and references therein.

An essential element of QFT description of oscillations is the neutrino
propagator. In electroweak theory a spontaneous symmetry breaking takes place in
a scalar sector that generates mass matrix for fermions. After diagonalization
of this matrix, a unitary mixing matrix arises in charged current vertex while
the neutral current vertex remains diagonal.

However, if to take into account self-energy contributions in propagator, the
picture is noticeably complicated algebraically due to appearance of
\(\gamma^{5}\) matrix and necessity of renormalization. These contributions may
originate from radiative corrections in perturbation theory, or through dynamic
symmetry breaking beyond the Standard Model, or through interaction with a
medium.

The renormalization of dressed fermion propagator with mixing between
generations was considered by a number of authors. The basic requirements for
renormalization were formulated in \cite{Aoki:1982ed} and have subsequently been
used in a number of papers. As a rule, the consideration is restricted by the
first corrections of perturbation theory.

One should point out the results of \cite{Kni12, Kni14PR}, where explicit
formulas for matrix propagator in all orders of perturbation theory were
derived. Similar results were obtained in \cite{Ben10}, where models with
dynamic generation of fermion masses were discussed. The used there inversion
procedure is non-trivial in presence of \(\gamma^{5}\), leads to non-obvious
renormalization procedure, see detailed study in \cite{Kni14PR}.

It is known that accounting of radiative corrections in propagator (and vertex)
leads to renormalization of quark or neutrino mixing matrix. This problem was
considered in various approaches and approximations in QFT framework, see, e.g.,
\cite{Denner:1990yz,Gambino:1998ec,Barroso:2000is,Kniehl:2006rc,Duret:2008st,
  Esp02}.

Earlier we investigated \cite{KL16} the eigenvalue problem for fermion
propagator with mixing between generations. The properties of the obtained
spectral representation \cite{KL16} allow to represent propagator \(G(p)\) for
fermion system as sum of single poles accompanied by orthogonal projectors. Due
to algebraic simplicity, such construction allows to renormalize matrix
propagator without referring to perturbation theory \cite{KL16}.

Here we discuss diagonalization procedure for matrix propagator based on the
eigenvalue problem. We found that this problem allows to write down explicit
form of similarity transformation, converting inverse propagator \(S(p)\) into a
diagonal form. After that the inversion procedure becomes obvious. The obtained
diagonalization of a dressed propagator allows to renormalize mixing matrix
without using of perturbation theory.

\section{Matrix propagator and eigenvalue problem}
\label{sec:matr-prop-eigenv}

\subsection{Dressed propagator and basis}
\label{sec:dress-prop-basis}

In case of \(n\) mixing fermions (quarks or leptons), the inverse propagator
with account of self-energy terms is
\begin{equation}\label{Sgen}
  \begin{lgathered}
    S=\hat{p} - M^{\diag} - \Sigma(p), \quad
    \Sigma(p) = A(p^{2}) + \hat{p} B(p^{2}) + \gamma^{5} C(p^{2}) + \hat{p}
    \gamma^{5} D(p^{2}),
  \end{lgathered}
\end{equation}
where coefficients, accompanying \(\gamma\)-matrices, are matrices of dimension
\(n\).

Below we will use the off-shell \(\gamma\)-matrix projectors
\begin{equation}\label{lam}
  \operP_{1,2}=\frac{1}{2} \Big( 1\pm \frac{\hat{p}}{W} \Big),
\end{equation}
where \(W=\sqrt{p^{2}}\) is invariant mass. In theory with parity violation it
is convenient to introduce the following set of \(\gamma\)-matrix operators with
simple algebraic properties
\begin{equation}\label{basis}
  \operP_{1},\quad
  \operP_{2},\quad
  \operP_{3}=\operP_{1}\gamma^{5},\quad
  \operP_{4}=\operP_{2}\gamma^{5},
\end{equation}
which will be used below as a basis. Matrix inverse propagator can be written as
an expansion on this basis
\begin{equation}\label{decomp}
  S(p) = G^{-1}(p)= \sum_{M=1}^{4} \operP_{M} S_{M}(W) .
\end{equation}

Below we will consider the simplest case, when coefficients in self-energy
\(\Sigma(p)\) are real functions. Furthermore, we will restrict ourselves to
case of \(\mathsf{CP}\)-conservation\footnote{Below a threshold this coincides
  with pseudo-hermitian condition \(S=\gamma^{0} S^{\dag} \gamma^{0}\).}, which
leads to the symmetry of matrix coefficients \eqref{Sgen}, see, e.g.,
\cite{Kni08}
\begin{equation}
  A^{\Tp}=A,\quad B^{\Tp}=B,\quad D^{\Tp}=D,\quad C^{\Tp}=-C .
\end{equation}
For expansion coefficients \eqref{decomp} this gives:
\begin{equation}\label{symm}
  (S_{1,2})^{\Tp}=S_{1,2}, \quad (S_{3})^{\Tp}=-S_{4}. 
\end{equation}
It is known that for regular matrix the diagonalization is carried out on the
base of eigenvalue problem. The inverse propagator \eqref{Sgen} is ``double''
matrix with two different sets of indices. But as we will see later, to
diagonalize such operator it is also sufficient to solve the eigenvalue problem.

\subsection{Diagonal form of propagator}
\label{sec:diag-form-prop}

In case of \(n\) mixing fermions, it is not quite obvious what the diagonal form
of propagator is. Let us consider free propagator with diagonal mass matrix and
write it down using the off-shell projectors \(\operP_{1,2}\):
\begin{multline}
  \label{diag0}
  S_{0}(p) = \hat{p} - M^{\diag} =\\
  =\operP_{1}
  \begin{pmatrix}
    &W-m_{1} &  & 0 & \\
    &  & \ddots & & \\
    & 0 & & W-m_{n}&
  \end{pmatrix} 
  +\operP_{2}
  \begin{pmatrix}
    &-W-m_{1} &  & 0 & \\
    &  & \ddots & & \\
    & 0 & & -W-m_{n}&
  \end{pmatrix}.
\end{multline}
One can see that here the diagonal elements \((\pm W -m_{i})\) are eigenvalues
of operator \(S_{0}\). Indeed, eigenprojector for \(S_{0}\) looks like
\begin{equation}\label{pro_{0}}
  \Pi_{i}^{(+)} = \mathcal{P}_{1} \pi^{i}, \qquad \quad
  \Pi_{i}^{(-)} = \mathcal{P}_{2} \pi^{i} ,
\end{equation}
where \(\pi^{i}\) is elementary \(n\times n\) projector
\begin{equation}\label{eq:pi-elem-proj}
  ( \pi^{i} )_{lm}= \delta_{il} \delta_{im}
\end{equation}
consisting of zeros and unit at \(i\)-th position on the diagonal. On the mass
shell \(\Pi_{i}^{(\pm)}\) give rise to solutions of Dirac equation with positive
(negative) energy. It is evident that projectors \eqref{pro_{0}} are solutions
of eigenvalue problem for bare inverse propagator \eqref{diag0}:
\begin{equation}
  S_{0} \Pi_{i}^{(\pm)} =  (\pm W -m_{i}) \Pi_{i}^{(\pm)} .
\end{equation}

Now consider the dressed matrix inverse propagator \eqref{Sgen}. In analogy with
\eqref{diag0} we should define
\begin{equation}
  \label{diag}
  \begin{lgathered}
    S^{\diag}(p) = \operP_{1}
    \begin{pmatrix}
      &\lambda_{1} &  & 0 & \\
      &  & \ddots & & \\
      & 0 & & \lambda_{n}&
    \end{pmatrix}
    + \operP_{2}
    \begin{pmatrix}
      &\lambda_{n+1}  &  & 0 & \\
      &  & \ddots & & \\
      & 0 & & \lambda_{2n}&
    \end{pmatrix} \equiv \operP_{1}\Lambda^{+} + \operP_{2} \Lambda^{-}
  \end{lgathered}
\end{equation}
as the diagonal form of matrix inverse propagator. Here \(\lambda_{i}(W)\) are
eigenvalues of inverse propagator. First \(n\) of them
\(\lambda_{1},\ldots,\lambda_{n}(W)\) correspond to positive energy solutions,
and \(\lambda_{n+1},\ldots,\lambda_{2n}\) to negative energy ones — cf. with
\eqref{diag0}. The diagonal form \eqref{diag} can be easily reduced to
frequently used form with unit kinetic matrix. However, the diagonal form of
propagator used by us is some part of general algebraic construction.

\subsection{Eigenvalue problem for matrix propagator}
\label{sec:eigenv-probl-matr}

Eigenvalue problem for operator of form \eqref{Sgen} was investigated in
\cite{KL16}. Here we give only necessary information for case of
\(\mathsf{CP}\)-conservation \eqref{symm}.

Note that it is more convenient to solve this problem in a matrix form, i.e. to
search not eigenvectors but eigenprojectors of operator \(S(p)\). As in case of
matrix \(n\times n\) of general form, it is necessary to solve two problems:
left and right ones
\begin{equation}
  \label{eigenLR}
  S \Pi = \lambda \Pi, \qquad \quad
  \Pi S= \lambda \Pi ,
\end{equation}
using decomposition of form \eqref{decomp} for both operator \(S\) and required
projector.

Assuming \(\mathsf{CP}\)-conservation, the problem of eigenprojector
construction is reduced to solution of homogeneous equation
\begin{equation}
  \label{homo} 
  \hat{O}\psi \equiv
  \big[(S_{2}-\lambda)S_{3}^{-1}(S_{1}-\lambda)-S_{4}\big] \psi =0 ,
\end{equation}
where \(\hat{O}\) is a matrix \(n\times n\) and \(\psi\) is a vector of
dimension \(n\). The eigenvalues \(\lambda_{i}\) are determined by
characteristic equation
\begin{equation}\label{det}
  \det \big[(S_{2}-\lambda)S_{3}^{-1}(S_{1}-\lambda)-S_{4}\big] = 0.
\end{equation}

The solution of left and right problems \eqref{eigenLR} can be written as
decomposition over basis \eqref{decomp} (see details in \cite{KL16}):
\begin{equation}
  \label{L_{R}}
  \Pi_{i} = \varepsilon_{i} \Big(
    \operP_{1}\cdot \psi_{i} {\psi}_{i}^{\Tp} -
    \operP_{2}\cdot \phi_{i} {\phi}_{i}^{\Tp} +
    \operP_{3}\cdot \psi_{i} {\phi}_{i}^{\Tp} -
    \operP_{4}\cdot \phi_{i} {\psi}_{i}^{\Tp}
  \Big).
\end{equation}
Here \(\varepsilon_{i} =\pm 1\) are energy sign, vectors \(\psi_{i}\) are
solutions of homogeneous equations
\begin{equation}
  \label{eq_{p}si}
  \hat{O}_{i} \psi_{i} \equiv \hat{O}(\lambda = \lambda_{i})\psi_{i}= 0,
\end{equation}
and vectors \(\phi_{i}\) are expressed through them by
\(\phi_{i} = S_{3}^{-1} (S_{1} - \lambda_{i}) \psi_{i}\).
Let us require matrices \eqref{L_{R}} to be orthogonal projectors
\begin{equation}
  \Pi_{i} \Pi_{k} = \delta_{ik}\Pi_{k},\quad
  i,k=1, \dots, 2n.
\end{equation}
It gives the orthogonality conditions
\begin{equation}
  \label{ortho}
  \varepsilon_{i}  \big( {\psi}_{i}^{\Tp}  \psi_{k}  -
    {\phi}_{i}^{\Tp} \phi_{k}  \big) = \delta_{ik}.
\end{equation}
For \(i\neq k\) this condition follows from homogeneous equations and for
\(i=k\) it defines normalization of \(\psi\).

The other necessary requirement for system of orthogonal projectors
\eqref{L_{R}} is the completeness condition
\begin{equation}
  \label{com}
  \sum^{2n}_{i=1} \Pi_{i} = 1 \equiv I_{4} I_{n} .
\end{equation}
The unit \(\gamma\)-matrix in our basis is \(I_{4} = \operP_{1}+ \operP_{2}\),
therefore the completeness condition is reduced to requirements on matrix
coefficients of eigenprojector \eqref{L_{R}}
\begin{equation}
  \label{com1}
  \begin{lgathered}
    \sum_{i=1}^{2n} \varepsilon_{i} \psi_{i} {\psi}_{i}^{\Tp} = I_{n},\quad
    \sum_{i=1}^{2n} \varepsilon_{i} \phi_{i} {\phi}_{i}^{\Tp} =-I_{n},\\
    \sum_{i=1}^{2n} \varepsilon_{i} \psi_{i} {\phi}_{i}^{\Tp} = 0_{n},\quad
    \sum_{i=1}^{2n} \varepsilon_{i} \phi_{i} {\psi}_{i}^{\Tp} = 0_{n} .
  \end{lgathered}
\end{equation}
As is shown in 
\ref{sec:prop}, the completeness condition follows from the orthogonality
conditions and vice verse.

\section{Matrix propagator diagonalization}
\label{sec:matr-prop-diag}

As a result of solving of the left and right eigenvalue problems the inverse
matrix propagator \(S\) can be written in form of spectral decomposition
\begin{equation}
  \label{spectr}
  S= \sum_{i=1}^{2n} \lambda_{i} \Pi_{i} \qquad \Rightarrow \qquad
  G= S^{-1} = \sum_{i=1}^{2n} \frac{1}{\lambda_{i} } \Pi_{i},
\end{equation}
where eigenvalues \(\lambda_{i}\) are determined by algebraic equation
\eqref{det} and eigenprojectors \(\Pi_{i}\) have form \eqref{L_{R}}. To
construct the projectors, it is necessary to solve the homogeneous equation
\eqref{eq_{p}si} for each \(i\).

To bring \(S\) to a diagonal form, let us construct matrices of dimension \(n\)
\begin{equation}
  \label{mats}
  \begin{split}
    \Psi_{+} &=
    \begin{pmatrix}
      \psi_{1} \dots \psi_{n}
    \end{pmatrix}, \qquad
    \Psi_{-} =
    \begin{pmatrix}
      \psi_{n+1} \dots \psi_{2n}
    \end{pmatrix},\\
    \Phi_{+} &=
    \begin{pmatrix}
      \phi_{1} \dots \phi_{n}
    \end{pmatrix}, \qquad
    \Phi_{-} =
    \begin{pmatrix}
      \phi_{n+1} \dots \phi_{2n}
    \end{pmatrix}
  \end{split}
\end{equation}
from vector-columns \(\psi_{i}\), \(\phi_{i}\) involved in the eigenprojector
\eqref{L_{R}}. In terms of these matrices the desired diagonalization looks as a
similarity transformation:
\begin{equation}
  \label{trans}
  S= T^{-1} S^{\,\diag}\, T ,
\end{equation}
where the diagonal form of propagator is defined by \eqref{diag} and operators
\(T\), \(T^{-1}\) (their properties can be found in \ref{sec:prop}) have form
\begin{equation}
  \label{TTCP}
  \begin{split}
    T= \operP_{1} ({\Psi}_{+}^{\Tp}+ \gamma^{5} {\Phi}_{+}^{\Tp}) +
    \operP_{2} ({\Phi}_{-}^{\Tp} + \gamma^{5} {\Psi}_{-}^{\Tp})    , \\
    T^{-1} = ({\Psi}_{+} - \gamma^{5} {\Phi}_{+}) \operP_{1} + ({\Phi}_{-} -
    \gamma^{5} {\Psi}_{-}) \operP_{2} .
  \end{split}
\end{equation}
By writing separately contributions of energy projectors \(\operP_{1}\) and
\(\operP_{2}\) in inverse propagator \eqref{trans}, one obtains
\begin{equation}
  \label{proCP}
  \begin{lgathered}
    S= ({\Psi}_{+} - \gamma^{5} {\Phi}_{+}) \operP_{1} \Lambda^{+}
    ({\Psi}_{+}^{\Tp}+ \gamma^{5} {\Phi}_{+}^{\Tp})
    + ({\Phi}_{-} - \gamma^{5}
    {\Psi}_{-}) \operP_{2} \Lambda^{-} ({\Phi}_{-}^{\Tp} + \gamma^{5}
    {\Psi}_{-}^{\Tp}).
  \end{lgathered}
\end{equation}
Now one can identify eigenprojectors \(\Pi_{i}\) \eqref{L_{R}}, which are
accompanied by factors \(\lambda_{i}\), see spectral decomposition
\eqref{spectr}
\begin{equation}
  \label{proj}
  \begin{aligned}
    \Pi_{i} &= ({\Psi}_{+} - \gamma^{5} {\Phi}_{+}) \mathcal{P}_{1} \pi^{i}
    ({\Psi}_{+}^{\Tp} + \gamma^{5} {\Phi}_{+}^{\Tp}), \\ 
    \Pi_{i+n} &=({\Phi}_{-} - \gamma^{5} {\Psi}_{-}) \mathcal{P}_{2} \pi^{i}
    ({\Phi}_{-}^{\Tp} + \gamma^{5} {\Psi}_{-}^{\Tp}),
  \end{aligned}
  \qquad i=1,\ldots,n.
\end{equation}
As a result, we see that the factorized form \eqref{trans} of inverse propagator
is equivalent to spectral representation \eqref{spectr}.

Propagator \(G(p)\) is obtained by inverting of \eqref{trans}
\begin{equation}
  \label{gfact}
  \begin{lgathered}
    G = T^{-1} G^{\,\diag}\, T 
    =T^{-1} \Big[ \operP_{1} \big(\Lambda^{+}\big)^{-1} +
      \operP_{2} \big(\Lambda^{-}\big)^{-1} \Big] T =  \\
    = ({\Psi}_{+} - \gamma^{5} {\Phi}_{+}) \operP_{1} \big(\Lambda^{+}\big)^{-1}
    ({\Psi}_{+}^{\Tp}+ \gamma^{5} {\Phi}_{+}^{\Tp}) 
    + ({\Phi}_{-} - \gamma^{5} {\Psi}_{-}) \operP_{2} \big(\Lambda^{-}\big)^{-1}
    ({\Phi}_{-}^{\Tp} + \gamma^{5} {\Psi}_{-}^{\Tp}) .
  \end{lgathered}
\end{equation}

\section{Mixing matrix modification}
\label{sec:mixing-matr-modif}

Let us consider an elementary block of some diagram: quark vertex of changed
current surrounded by renormaziled dressed propagators.
\begin{center}
  \includegraphics[scale=0.9]{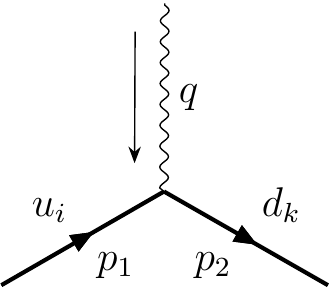}
\end{center}

The renormalization of fermion mixing matrix renormalization was discussed
earlier (see
\cite{Denner:1990yz,Gambino:1998ec,Barroso:2000is,Kniehl:2006rc,Duret:2008st,
  Esp02} and references therein), the consideration is usually based on one-loop
corrections. The main issues here are related with gauge invariance and
influence of renormalization scheme.

We will restrict ourselves by a simple example, when bare left vertex is
surrounded by dressed propagators in factorized form \eqref{gfact}. The main
goal is to investigate algebraic properties of proposed diagonalization. Thus,
consider the expression
\begin{equation}
  \label{vert}
  V^{\mu}= G^{(d)}(p_{2})\cdot K \gamma^{\mu} (1-\gamma^{5}) \cdot G^{(u)}(p_{1}).
\end{equation}
Here \(K\) is a bare Cabibbo-Kobayashi-Maskawa (CKM) matrix, appeared after
spontaneous symmetry breaking, \(G(p)\) is dressed matrix propagator
\eqref{gfact}.

We are interested in mixing matrix modification, so we consider only the factors
between poles in \eqref{vert}:
\begin{equation*}
  \begin{lgathered}
    \Gamma^{\mu} \equiv \big\{ G^{\diag} T\big\}_{p_{2}}^{(d)} \cdot K
    \gamma^{\mu} (1-\gamma^{5})\cdot \big\{ T^{-1}G^{\diag}\big\}_{p_{1}}^{(u)}
    = \\
    = \Big\{ \Big[ \operP_{1} (\Lambda^{+})^{-1} + \operP_{2}
    (\Lambda^{-})^{-1} \Big] T\Big\}_{p_{2}}^{(d)} 
    \cdot K \gamma^{\mu} (1-\gamma^{5})
    \cdot \Big\{ T^{-1}\Big[ \operP_{1} (\Lambda^{+})^{-1} +
    \operP_{2} (\Lambda^{-})^{-1} \Big] \Big\}_{p_{1}}^{(u)}.
  \end{lgathered}
\end{equation*}
The left projector in the vertex turns \(\gamma^{5}\) in operators \(T\),
\(T^{-1}\) into unit but nevertheless they still contain \(\gamma\)-matrix
projectors \(\operP_{1}\), \(\operP_{2}\). Let us examine contributions of poles
with positive energy
\begin{equation}
  \label{vert2}
  \begin{aligned}
    \Gamma_{++}^{\mu} & \equiv \big\{(\Lambda^{+})^{-1} \operP_{1}
    ({\Psi}_{+}^{\Tp}+ \gamma^{5} {\Phi}_{+}^{\Tp}) \big\}_{p_{2}}^{(d)}
    \cdot K \gamma^{\mu} (1-\gamma^{5}) 
    \cdot\big\{ ({\Psi}_{+} - \gamma^{5} {\Phi}_{+})\operP_{1}
    (\Lambda^{+})^{-1} \big\}_{p_{1}}^{(u)}  = \\
    &= \big\{(\Lambda^{+})^{-1} \operP_{1}
    ({\Psi}_{+}^{\Tp}+{\Phi}_{+}^{\Tp}) \big\}_{p_{2}}^{(d)}
    \cdot K \gamma^{\mu} (1-\gamma^{5}) 
    \cdot\big\{ ({\Psi}_{+} + {\Phi}_{+})\operP_{1}
    (\Lambda^{+})^{-1} \big\}_{p_{1}}^{(u)}.
  \end{aligned}
\end{equation}
So, one can see the off-shell CKM matrix modification:
\begin{equation}
  \label{KPR}
  K \to K^{\prime} =
  \big( {\Psi}_{+}^{\Tp}+{\Phi}_{+}^{\Tp} \big)_{p_{2}}^{(d)}
  \cdot K \cdot
  \big( {\Psi}_{+}+{\Phi}_{+} \big)_{p_{1}}^{(u)} .
\end{equation}
The mixing matrix is modified by matrices \(({\Psi}_{+}+{\Phi}_{+})\),
\(({\Psi}_{+}+{\Phi}_{+})^{\Tp}\), dependent on different momenta \(p_{1}\),
\(p_{2}\).

Let us check whether \(K'\) is a unitary matrix
\begin{equation}
  \label{KK}
  \begin{lgathered}
    K^{\prime}\, K^{\prime\,\dag} =
    \big({\Psi}_{+}+{\Phi}_{+}\big)^{\Tp}_{p_{2}} K
    \big({\Psi}_{+}+{\Phi}_{+}\big)_{p_{1}}
    \cdot\big({\Psi}_{+}+{\Phi}_{+}\big)^{\dag}_{p_{1}} K^{\dag}
    \big({\Psi}_{+}+{\Phi}_{+}\big)^{*}_{p_{2}}
  \end{lgathered}
\end{equation}
If propagators are real functions, then matrices \(\Psi_{\pm}\),
\(\Phi_{\pm}\) should be taken as real. Consider appeared in \eqref{KK} matrix
depending on \(p_{1}\):
\begin{equation}
  X(p_{1}) = \big({\Psi}_{+}+{\Phi}_{+}\big)
  \big({\Psi}_{+}+{\Phi}_{+}\big)^{\Tp}.
\end{equation}
Using the completeness condition \eqref{com_{m}} one can see that the off-shell
matrix \(X(p_{1})\) is not a unit one, therefore off-shell matrix \(K'\) cannot
be unitary.

If we want to consider the case of real fermions, we need to keep only one pole
in initial and final propagators. Corresponding residue, in analogy with LSZ
repice \cite{Lehmann:1954rq}, will give a transition amplitude.

If outgoing line in diagram is on mass shell, then \(W_{2} \to m^{(d)}_{k}\) and
\(\lambda_{k}\to W_{2}-m_{k}^{(d)}\) in propagator — see \eqref{vert2}. As for
diagonal matrix of eigenvalues, one should keep only pole term \(1/\lambda_{k}\)
\begin{equation}
  \label{ms}
  (\Lambda^{+})^{-1}_{p_{2}} = \begin{pmatrix}
    &1/\lambda_{1} &  & 0 & \\
    &  & \ddots & & \\
    & 0 & & 1/\lambda_{n}&
  \end{pmatrix}_{p_{2}}^{(d)}
  \to \frac{1}{\lambda_{k}}\pi^{k}\quad \text{at}\; W_{2} \to m^{(d)}_{k}.
\end{equation}
Next, in the accordance with LSZ prescription, we need to calculate residue at
the pole. This means that in matrix of inverse eigenvalues, involving in
\eqref{vert2}, the substitution
\begin{equation}
  (\Lambda^{+})^{-1}_{p_{2}}  
  \; \to \; \Delta^{k}  \equiv
  \delta(W_{2} - m^{(d)}_{k}) \pi^{k}
\end{equation}
should be done. Properties of vectors \(\psi_{i}\), \(\phi_{i}\) in the
eigenprojectors \eqref{L_{R}} were studied in \cite{KL16}. In terms of
  \(n\times n\) matrices \eqref{mats}, the renormalization conditions give:
\begin{equation}
  \label{eq:1}
  \Psi_{+}\to \pi^{k},\quad \Phi_{+}\to 0 \quad \text{at}\; W\to m_{k}.
\end{equation}
Then one can see, that solution matrix ``under observation'' of \(\Delta^{k}\)
turns into unit
\begin{equation}
  \Delta^{k} \cdot \big({\Psi}_{+}+{\Phi}_{+}\big)^{\Tp} =
  \Delta^{k} \cdot   \pi_{k} = \Delta^{k}  \cdot I_{n} .
\end{equation}
Similar simplification occurs for incoming quark in diagram.

As a result, we see that for fermions on mass-shell \(K'=K\), i.e. accounting
radiative corrections in fermion propagator does not change bare mixing matrix.

\section{Conclusion}
\label{sec:conclusion}

We have considered diagonalization of dressed fermion propagator with parity
violation and mixing between generations. It turned out that solution of
eigenvalue problem \cite{KL16} for inverse propagator allows to bring propagator
to a diagonal form. In this case the transformation operator \eqref{TTCP} is
``double'' matrix having generation and flavor indexes. As for the diagonal form
of propagator, we propose to use the most natural definition \eqref{diag} based
on the eigenvalue problem.

Let us recall that CKM matrix is appeared at tree level after diagonalization of
a mass matrix with use of polar decomposition.  This variant of diagonalization
was applied also to a fermion mixing problem \cite{Duret:2008st} at loop level,
though not for the very general case.  We suppose that comparison of different
variants of propagator diagonalization should be interesting.

The obtained diagonalization in principle allows to investigate modification of
mixing matrix in charged current vertex without referring to perturbation
theory.  Here we constrained ourselves by simple example involving bare left
vertex surrounded by dressed propagators. We think that the suggested form of
propagator can serve a useful tool for futher studing of fermion mixing problem
in more complicated situations.

\section{References}
\bibliography{diag.bib}
\bibliographystyle{unsrtnat}
\label{sec:references}

\newpage
\appendix	
\section{Properties of \(T\), \(T^{-1}\) operators}
\label{sec:prop}

The orthogonality property \eqref{ortho} can be restated in terms of matrices
\eqref{mats} constructed from vector-solutions.
\begin{equation}
  \label{ort_{m}}
  \begin{split}
    {\Psi}_{+}^{\Tp} {\Psi}_{+} - {\Phi}_{+}^{\Tp}{\Phi}_{+}= I_{n}, \quad
    {\Phi}_{-}^{\Tp} {\Phi}_{-} - {\Psi}_{-}^{\Tp}{\Psi}_{-}=I_{n}, \\
    {\Psi}_{+}^{\Tp} {\Psi}_{-} - {\Phi}_{+}^{\Tp}{\Phi}_{-}=0_{n}, \quad
    {\Psi}_{-}^{\Tp} {\Psi}_{+} - {\Phi}_{-}^{\Tp}{\Phi}_{+}=0_{n}.
  \end{split}
\end{equation}
To verify let us calculate matrix element of matrices product:
\begin{equation}
   ({\Psi}_{+}^{\Tp} {\Psi}_{+})_{ij}= ({\Psi}_{+}^{\Tp})_{ik}
   ({\Psi}_{+})_{kj}=
   ({\Psi}_{+})_{ki} ({\Psi}_{+})_{kj}
   =({\psi}_{i})_{k} ({\psi}_{j})_{k}= ({\psi}_{i})^{T} ({\psi}_{j})
\end{equation}
In last expression we returned to vectors \({\psi}_{i}\), the solutions of
homogeneous equations. We see that matrix conditions \eqref{ort_{m}} are
equivalent to \eqref{ortho}.

The completeness conditions \eqref{com1} in terms of matrices become:
\begin{equation}
  \label{com_{m}}
  \begin{split}
    {\Psi}_{+} {\Psi}_{+}^{\Tp} - {\Psi}_{-}{\Psi}_{-}^{\Tp} = I_{n}, \quad
    {\Phi}_{-}{\Phi}_{-}^{\Tp} - {\Phi}_{+}{\Phi}_{+}^{\Tp}=I_{n}, \\
    {\Psi}_{+} {\Phi}_{+}^{\Tp} - {\Psi}_{-}{\Phi}_{-}^{\Tp} = 0_{n}, \quad
    {\Phi}_{-}{\Psi}_{-}^{\Tp} - {\Phi}_{+}{\Psi}_{+}^{\Tp}=0_{n} .
  \end{split}
\end{equation}

Let us check that operators \eqref{TTCP} are indeed inverse ones. Multiply them
using properties of \(\gamma\)-matrix basis \eqref{basis}
\begin{equation*}
  \begin{lgathered}
    T\cdot T^{-1} = \Big[ \operP_{1} \big({\Psi}_{+}^{\Tp} + \gamma^{5}
    {\Phi}_{+}^{\Tp}\big) + \operP_{2}\big({\Phi}_{-}^{\Tp} + \gamma^{5}
    {\Psi}_{-}^{\Tp}\big) \Big] 
    \cdot \Big[ \big({\Psi}_{+} - \gamma^{5} {\Phi}_{+}\big) \operP_{1} +
    \big({\Phi}_{-} - \gamma^{5} {\Psi}_{-}\big) \operP_{2} \Big] =  \\
    =\operP_{1} \big({\Psi}_{+}^{\Tp} {\Psi}_{+} -
    {\Phi}_{+}^{\Tp}{\Phi}_{+}\big) +
    \operP_{2} \big({\Phi}_{-}^{\Tp} {\Phi}_{-} - {\Psi}_{-}^{\Tp}{\Psi}_{-}\big) 
    - \operP_{3} \big({\Psi}_{+}^{\Tp} {\Psi}_{-} -
    {\Phi}_{+}^{\Tp}{\Phi}_{-}\big) - \operP_{4} \big({\Phi}_{-}^{\Tp}
    {\Phi}_{+} -
    {\Psi}_{-}^{\Tp}{\Psi}_{+}\big) =\\
    = \operP_{1} I_{n} + \operP_{2} I_{n} = I_{4} I_{n} .
  \end{lgathered}
\end{equation*}
In last step we used matrix orthogonality conditions \eqref{ort_{m}}.

On the other hand, from the associativity of matrix multiplication follows that
\(T^{-1}\cdot T=1\). Writing down corresponding calculations
\begin{equation*}
  \begin{lgathered}
    T^{-1} \cdot T = \Big[ \big({\Psi}_{+} - \gamma^{5} {\Phi}_{+}\big) \operP_{1} +
    \big({\Phi}_{-} - \gamma^{5} {\Psi}_{-}\big) \operP_{2} \Big] 
    \cdot
    \Big[  \operP_{1} \big({\Psi}_{+}^{\Tp} + \gamma^{5} {\Phi}_{+}^{\Tp}\big) +
    \operP_{2}\big({\Phi}_{-}^{\Tp} + \gamma^{5} {\Psi}_{-}^{\Tp}\big)  \Big] = \\
    = \operP_{1} \big(\Psi_{+} {\Psi}_{+}^{\Tp} - \Psi_{-} {\Psi}_{-}^{\Tp}\big) +
    \operP_{2} \big(\Phi_{-} {\Phi}_{-}^{\Tp} - \Phi_{+} {\Phi}_{+}^{\Tp}\big)
    + \operP_{3} \big(\Psi_{+} {\Phi}_{+}^{\Tp} - \Psi_{-} {\Phi}_{-}^{\Tp}\big) +
    \operP_{4} \big(- \Phi_{+} {\Psi}_{+}^{\Tp} + \Phi_{-} {\Psi}_{-}^{\Tp}\big)
    =\\
    = I_{4} I_{n} .
  \end{lgathered}
\end{equation*}
give us the completeness conditions \eqref{com_{m}}. Thus, one sees that from
orthogonality condition \eqref{ortho} of eigenprojectors follows the
completeness condition \eqref{com} and vice verse.

\end{document}